\documentclass[11pt]{article}

\usepackage[dvipsnames]{xcolor}
\usepackage{hyperref}
\usepackage{amsmath}
\usepackage{cleveref}
\usepackage{adjustbox}
\usepackage{multirow}
\usepackage{multicol}

\newcommand\myshade{90}
\colorlet{mylinkcolor}{NavyBlue}
\colorlet{mycitecolor}{Aquamarine}
\colorlet{myurlcolor}{Aquamarine}

\hypersetup{
  linkcolor  = mylinkcolor!\myshade!black,
  citecolor  = mycitecolor!\myshade!black,
  urlcolor   = myurlcolor!\myshade!black,
  colorlinks = true,
}

\usepackage{graphicx}
\usepackage{grffile}
\usepackage{subcaption}
\usepackage[percent]{overpic}
\usepackage{booktabs} 
\usepackage{multirow}

\usepackage{url}            
\usepackage{amsfonts}       
\usepackage{color}
\usepackage{wrapfig}

\usepackage{tikz}

\usepackage[margin=2.6cm]{geometry}
\usepackage{mathtools}
\usepackage{amssymb}
\usepackage{nicefrac}
\usepackage{parskip}
\usepackage[numbers]{natbib}
\usepackage{authblk}

\usepackage{microtype}

\title{Cell image classification: a comparative overview}

\author[1,2]{Mohammad Shifat-E-Rabbi}
\author[1,3]{Xuwang Yin}
\author[1,2]{Cailey Elizabeth Fitzgerald}
\author[1,2,3,*]{Gustavo K. Rohde}
\affil[1]{Imaging and Data Science Lab, Charlottesville, VA-22903, USA}
\affil[2]{Department of Biomedical Engineering, University of Virginia, Charlottesville, VA-22903, USA}
\affil[3]{Department of Electrical \& Computer Engineering, University of Virginia, Charlottesville, VA-22903, USA}

\affil[*]{gustavo@virginia.edu}

\date{}

\begin{document}
\maketitle

\begin{abstract}
Cell image classification methods are currently being used in numerous applications in cell biology and medicine. Applications include understanding the effects of genes and drugs in screening experiments, understanding the role and subcellular localization of different proteins, as well as diagnosis and prognosis of cancer from images acquired using cytological and histological techniques. We review three different approaches for cell image classification: numerical feature extraction, end to end classification with neural networks, and transport-based morphometry. In addition, we provide comparisons on four different cell imaging datasets to highlight the relative strength of each method\footnote{\textcopyright{ \it{Cytometry Part A}} (2020) 97 (4), 347-362. Permission from the journal must be obtained for all uses.}.
\end{abstract}

\section{Introduction}
Interpretation of images of cells has always played important roles in science and medicine. From their discovery in 1665, observation of the spatiotemporal characteristics of cells through microscopic technology has enabled us to better understand the structure of living cells, as well as how they perform certain functions \cite{hooke1961micrographia, mazzarello1999unifying}. In addition, scientists have long used microscopes to evaluate the efficacy of different compounds for drug development \cite{perlman2004multidimensional,scheeder2018machine, xu2008, basu2014detecting, ohnuki2010, Caicedo2017, murali2018}. In medicine, as another example, the observation of cell morphology has long been used to discern malignancy in cancer cells \cite{simon1998automated, kantara2015methods, moallemi1991classifying, lee2003adaptive}.

Cells are known to exhibit complex phenotypes such as differences in shape, gene expression, subcellular protein localization, and other qualities. In addition, cell cultures, tissues, and organs are known to exhibit complex heterogeneity of phenotypes. The combination of intricate phenotype differences together with their heterogeneous responses to different conditions (e.g. diseases) has made decoding biological processes an increasingly complex task. Thus computational approaches for analyzing images of cells have been used increasingly to aid in the task of decoding the complexity of biological processes. A common task useful in many practical situations is determining the category of a given cell or set of cells: a task known as cell classification. 

Automated cell classification via computational analysis of images of cells have found numerous applications in science, technology, and medicine. Scientists have long used cell classification methods to determine whether a particular drug has affected a given culture of cells in the desired manner \cite{perlman2004multidimensional,scheeder2018machine, xu2008, basu2014detecting, ohnuki2010, Caicedo2017, murali2018}. In pathology, an increasing number of researchers are exploring methods to automatically detect cancer based on classification of images of cells \cite{simon1998automated, kantara2015methods, moallemi1991classifying, lee2003adaptive}. As another example, geneticists have also resorted to automated cell classification in attempts to study gene silencing mechanisms \cite{conrad2010automated,jones2009scoring}. These and other applications are reviewed in the next section.

In the context of the cell images, there are three broad categories of image classification algorithms that are used currently and reviewed in this article: feature extraction and machine learning, neural networks, and transport-based morphometry.
\begin{itemize}
    \item Feature engineering has been used for decades, by means of both manual and automated feature extraction \cite{prewitt1966analysis,ponomarev2014ana,hsieh2009hep, wiliem2013classification,ersoy2012hep, nosaka2014hep,manivannan2014hep,ko2011cell, tai2011blood, hamilton2007fast,kothari2013histological}. Features may be hand-picked by experts to distinguish data between classes based on prior knowledge, or existing feature extraction software can compute thousands of data features for classification \cite{boland2001neural, shamir2008}. Once numerical features are extracted, statistical regression methods for classification (e.g. linear discriminant analysis, random forests, neural networks) are then employed to build a cell image classification system.
    
    \item Artificial neural networks, especially deep learning such as convolutional neural networks (CNNs), have emerged as the leading end-to-end classification systems \cite{lecun2015,yusoff2010performance,gao2017hep,phan2016transfer,gao2014hep}. CNNs bypass the feature extraction process and instead utilize a number of convolutional layers connecting the input data to the rest of the neural network. Their performance has surpassed previous benchmarks for some tasks, usually in domains where a large amount of data relative to the problem complexity is available for training.

    \item Transport-based morphometry (TBM) is a technique for modeling and feature extraction that utilizes the mathematics of optimal transport \cite{wang2013linear, basu2014detecting}. Transport-based methods have been used successfully in biomedical data and image analysis tasks, such as statistical modeling and inverse problems \cite{kolouri2017}. The data transformation method is constructed by comparing data vectors with respect to both their functional values, such as signal or pixel intensities and their location, such as time or pixel coordinates. Unlike linear modeling techniques which compare values at fixed coordinates, the transport-based methods are rooted in the physics of biological processes which are governed by the continuity equation. 
\end{itemize} 

 The goal of this paper is to present a non-exhaustive comparative overview of cell image classification methods. We aim to review the three categories explained above in more detail, as well as provide a comparison of how they work in practice using four publicly available datasets.  
 Source code implementing methods reviewed and compared here is available at https://github.com/rohdelab/cell-image-classification.

This paper is organized as follows. In Section \ref{secApp}, different applications of cell image classification are explained. The three main classification methods are briefly reviewed in \ref{secAvl}, and more extensively described in Appendix A. The description of experimental (computational) methods is available in section \ref{secExp} with guidelines for using the source code in Appendix B. The results are presented in section \ref{secRes} with discussions in \ref{secDiscnConcl}.

\section{Applications}
\label{secApp}
Cellular mechanisms underlie every action and development in the body and the natural world at large. Cellular dynamics, in part, determine the body's response to drugs and treatment, disease progression, and contain genetic information which can elucidate why abnormalities occur in certain individuals and not others. Because of this, cell image classification has many important applications including drug discovery, digital pathology, genetic screening, and cell biology, among others. 

\subsection*{Drug discovery}
The drug discovery process involves many steps from identification of targeted symptoms and disease, creation of chemical compounds designed to effect changes in regions causing the underlying disease, and execution of ex-vivo and finally in-vivo experiments. One limiting factor that affects the drug discovery pipeline is the decision of which chemical compounds should be allocated experimental resources. There may be many compounds which have the potential to be effective, but a limitation in resources affects their ability to be tested. Scientists carry out ex-vivo experiments to identify the efficacy of potential drugs, and image classification is one way to evaluate which drugs perform the best.

Cell phenotype classification methods have been used many times to explore mechanisms of action, target efficacy, and toxicity of drugs \cite{scheeder2018machine, ljosa2013comparison, xu2008}. For example, Loo~\emph{et al.}\cite{loo2007image} profiled responses of different drugs using a supervised cell image classification with support vector machines to separate drug-treated and untreated human cancer cells using 296 predetermined phenotypical features. Ljosa~\emph{et al.}\cite{ljosa2013comparison} used both supervised and unsupervised classification algorithms with support vector machines and Gaussian mixture models, respectively, to differentiate mode of actions, e.g., Kinase inhibition, DNA replication, cholesterol regulation, etc. of a compendium of drugs e.g., alsterpaullone, camptothecin, and mevinolin, among others. The classification was performed in the space of 453 predetermined image features of breast cancer cells that were extracted using CellProfiler \cite{carpenter2006cellprofiler}.

Through high-throughput experimental methods, scientists are able to process large amounts of data necessary to identify which drugs to test further. Image-based cell classification is one way to test drug efficacy prior to in-vivo experiments, which are both costly and time-consuming. Researchers can treat specific cells with different drugs and examine the drug's effect by profiling cellular perturbations and comparing across drug treatment conditions \cite{perlman2004multidimensional, scheeder2018machine, basu2014detecting, ohnuki2010}. Classification methods allow scientists to discern whether differences in cell phenotypes arise as a result of drug application. In this way, scientists may be able to identify which drugs have the greatest potential to alleviate targeted symptoms and conditions prior to in-vivo experiments. A consequence of this benefit is that through computational resources, many more drugs can reach the ex-vivo stage for assessment prior to selection of which drugs to test in-vivo \cite{Caicedo2017, murali2018}.

\subsection*{Digital pathology}
One domain in which cellular classification techniques are already being implemented is diagnosis aided digital pathology. Through the combination of biopsy and imaging modalities, physicians can capture and examine a patient's cells from various regions of their body. Cellular imaging has become an important diagnostic tool, and advances in imaging techniques have resulted in an abundance of data; clinical decision support systems have arisen as a way to harness that data. Histological specimens are typically analyzed by trained experts, but there is inter-observer variability that can result in varying diagnoses \cite{comaniciu1999}. Computer-aided decision support systems increase the objectivity and reproducibility of diagnoses by relying on consistent algorithmic rules for  image-based classification. These algorithms also provide a way to grade and quantitatively assess the progression of a disease, a task which has previously been assessed via qualitative measures. Rojo~\emph{et al.}\cite{Garcia2006} have identified 31 commercially available digital slide systems, ten of which are equipped with technology to detect abnormalities in histological specimens. 

Advances in image classification have led to the first machine learning-based FDA-approved clinical decision support systems \cite{marr2017,  christiansen2018}. A typical flowchart of steps taken by a clinical decision support system, specifically in relation to histological tasks, include preprocessing, segmentation, feature extraction, dimension reduction, disease detection and classification, and post-processing and assessment \cite{he2012}. Cell image classification methods thus play important roles in diagnosis, and have increased cellular image understanding as well as boosted predictive analysis technologies \cite{yang2018, finkbeiner2015}. For example, Guillaud~\emph{et al.}\cite{guillaud2004quantitative} proposed a cervical cancer diagnosis procedure employing a supervised classification method using linear discriminant analysis. Numerical features of cellular images was used to classify cervical cancers into different grades: normal cervix, koilocytosis, and three stages of cervical intraepithelial neoplasia. Petushi~\emph{et al.}\cite{petushi2006large} reported a breast cancer diagnosis technique using a decision tree classifier with textural features of Hematoxylin and Eosin stained histology images to separate invasive breast carcinoma into three histologic grades.

One case for which cellular imaging is used to determine disease state and early intervention is in patients with Barrett's esophagus. Barrett's esophagus is a pre-malignant condition that predisposes patients to the development and progression of esophageal adenocarcinoma \cite{odze2006}. The condition can be detected by assessing tissues collected via endoscopic biopsy for levels of intestinal dysplasia \cite{vieth2004}, but a standard white-light examination of four-quadrant biopsy has been shown to miss neoplasia in 57\% of cases \cite{curvers2008}. Because morphological changes (i.e. increasing grades of epithelial dysplasia) indicate the carcinogenic progression of Barrett's mucosa, high-resolution imaging and cellular classification algorithms are important diagnostic tools \cite{polkowski1998}. In one study, a linear discriminant analysis-based classification algorithm on image features was used to increase diagnostic accuracy to 87\% \cite{muldoon2010}. 

\subsection*{Genetic screening}
Cell image classification techniques have extended the ability of researchers to investigate cellular genetic mechanisms. RNA interference (RNAi) is a term used to describe a mechanism which uses a cell's own DNA  sequence of a gene to \textit{silence} it, or reduce its functionality, and has emerged as a powerful tool to assess gene function \cite{Hannon2002}. When the normal function of a gene is required for a given cellular mechanism, knockdown of that gene may lead to a phenotype which is detectable via assay and associated imaging techniques \cite{Mohr2010}. Genetic screening via RNAi has already led to critical new insights for a number of pathologies and processes including infectious disease, cancer, aging, and drug discovery, among others \cite{Nijwening2010, Minois2010, Kassner2008, Cherry2009, Wolters2008}.

RNAi has increasingly been utilized in the context of cultured cells to perform large-scale genomic studies to identify multiple genes in a functional pathway. Because fluorescent microscopy allows one to visualize labeled proteins and their associated changes across conditions, high-throughput image classification tools can play an important role in helping researchers determine gene functionality and uncover previously uncharacterized phenotypes \cite{Mohr2012, conrad2010automated, jones2009scoring}. Supervised classification methods employing support vector machines used with statistical features of segmented cell images, e.g., shape, texture, homogeneity, brightness, etc. was suggested in Conrad~\emph{et al.}\cite{conrad2010automated} for separating different cellular morphologies in some RNAi screening applications. Hierarchical decision tree based classification has been used to separate different small interfering ribonucleic acid phenotypes to understand mechanisms of cell movement \cite{vitorino2008modular}.

\subsection*{Cell Biology}
Cellular level understanding of physiological processes can provide better insights into the mechanisms of tissue and organ systems. The use of computerized algorithms to aid in understanding biological processes is not new. The practice of karyotyping -- the process of describing the number and appearance of chromosomes in a eukaryotic cell in order to identify genomic defects -- has relied on automated procedures for over fifty years \cite{Gilbert1966, CASTLEMAN1976153}. Understanding the specific structure and characteristics of an organism's genetic make-up can provide insights into developmental status. Eliminating the human operator allows for not only the acceleration of the tedious process of finding cells in mitosis and arranging the chromosomal images into karyograms but also the quantization and classification of chromosomal features \cite{Danuser2011}. There are numerous other applications of computerized cell image classification techniques in cell and systems biology.

Cell signaling mechanisms regulate all cellular activities and thus can influence molecular-level physiology. Researchers have used Gaussian mixture models to detect subcellular particles from epifluorescence microscopic images and understand fusion and separation events, as well as signal transduction mechanisms of cell-surface receptors \cite{JAQAMAN2011593}. Additionally, scientists have classified cell images of different treatment conditions to recognize signaling regulation of morphological properties and protrusion and adhesion mechanisms of cells, as well as to monitor the temporal and hierarchical relationships among different signaling networks \cite{bakal2007quantitative}. In this application, supervised classification algorithms employing both linear discriminant analysis and neural networks were used with 154 numerical features of segmented cell images.

 
Automated cell image recognition techniques have also been used to discover important processes related to cell division, such as how chromosomes congress or segregate during mitosis \cite{GARDNER2008894}, the manner in which microtubules are organized in vertebrate meiotic spindles, and the process by which the bipolarity of the spindle is maintained throughout the cell division \cite{Yang2007}. Several works have employed computerized image detection techniques to recognize migratory patterns of cells explaining mechanisms of embryonic development, wound healing, and angiogenesis \cite{CHEN200965, SPRAGUE20033529, lewandowska1979attempt}. 

Boland~\emph{et al.}\cite{boland2001neural} classified subcellular protein localization patterns by training a neural network used with some predetermined numerical features of segmented HeLa cell images to understand the structures and functions of subcellular proteins. Monitoring the molecular architecture of every cell and analyzing tens of thousands of subcellular biomolecular markers is impractical for human operators \cite{CHEN200965, Keller1065, PONTI20053456}. Subcellular protein localization patterns provide information relating to the sequence, structure, and function of those proteins clarifying cell functions under different circumstances \cite{murphy2000towards, boland2001neural}. 

One exciting application of cell image classification is the recognition of biological processes which are invisible to the human eye. Hidden processes can be paired with measurable phenotypes by the use of mathematical models, and these processes can thereby be detected in association with any phenotypical change. This approach was first taken to understand the mechanism of kinetochore positioning in the metaphase spindle of yeast. Despite the unresolvable sizes of kinetochores, their movements were reliably detected in this study \cite{SPRAGUE20033529}.

\section{Overview of image classification methods}
\label{secAvl}
We have grouped the classification methods for distinguishing images of cells into three main categories: numerical feature engineering, neural networks, and transport-based morphometry. Here we provide a brief overview of each category. Rather than being exhaustive in the citation, we instead focus on describing broad trends, while selecting a few exemplary papers in each category to describe more carefully. The discussion in this section is kept at the overview level. Appendix A contains a more detailed explanation, including mathematical descriptions, of the methods.

\subsection*{Numerical feature engineering}
Numerical feature engineering (NFE) methods extract pre-determined features from segmented cells and use them to classify images with a statistical regression-based classifier \cite{ponomarev2014ana,murphy2000towards}. Cell images can be classified in their raw forms (e.g., pixel values) or can be summarized into some internal representation (or set of features) that may bear better discriminating information. A learning subsystem e.g., a classifier can then learn these features and differentiate images more effectively. 

\begin{figure*}
    \centering
    \includegraphics[width=17.5cm]{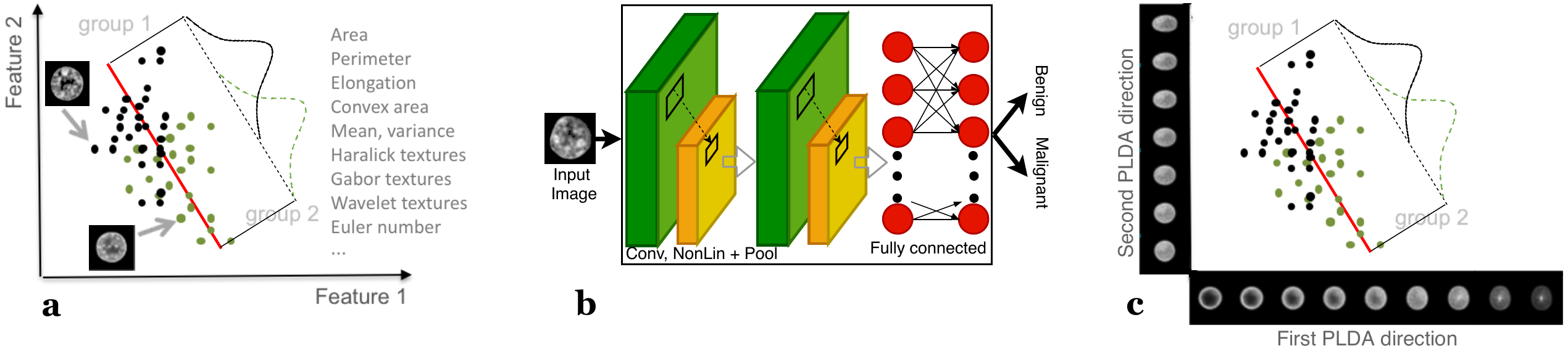}
    \caption{Cell image classification methods can be grouped into three broad categories: (A) numerical feature engineering (NFE), (B) neural networks (NN), and (C) transport-based morphometry (TBM).}
    \label{figmethod}
\end{figure*}

To better understand the concept of feature extraction in image classification, let us refer to Fig.~\ref{figmethod}(a). The goal here is to classify two groups of cell images. A feature engineering method represents each of the image samples (from both groups) in terms of a predefined feature set. Each point in the scatter plot denotes a feature-space representation of one particular cell image from either of two groups. A classifier can then be trained in this feature space to effectively differentiate two image groups. Appendix A contains a more detailed explanation of numerical feature engineering methods.

Numerical features-based cell fluorescence pattern classification has been used many times before \cite{ponomarev2014ana,hsieh2009hep, wiliem2013classification,ersoy2012hep, nosaka2014hep,manivannan2014hep} to detect different antinuclear auto-antibodies relevant to systemic autoimmune diseases. Numerical features have been used in other examples \cite{theera2007morphological,ko2011cell, tai2011blood, theera2005white} to count different types of white blood cells in bone marrow and thereby help in the diagnosis of certain diseases, e.g., leukemia, acquired immunodeficiency syndrome, cancers, etc. Numerical feature-based cell image classification has also been used to understand the sequence, structure, and function of different subcellular proteins \cite{hamilton2007fast, murphy2000towards}; and to differentiate histological images for automatic diagnosis of cancer \cite{kothari2013histological}. Engineers have experimented with a large number of image features: area, perimeter, elongation, convex area, mean, variance, Harlick/Gabor/wavelet textures, Euler numbers, and many more. Ideally, extracted features should be selective of aspects of images that are more relevant to discrimination. But the best discriminating features vary across datasets. Selection of features is an open problem of feature engineering methods. 

To evaluate performance of the numerical feature engineering category of methods, we used image features extracted by Wnd-chrm \cite{orlov2008wnd}. Performance metrics were obtained via implementation of existing classification algorithms from the scikit-learn package \cite{scikit-learn}: random forests (RF), k-nearest neighbors (k-NN), linear support vectors machines (SVM-l), logistic regression (LR), linear discriminant analysis (LDA), kernel support vector machines with radial basis function (SVM-k), and penalized linear discriminant analysis (PLDA). Appendix B has more information regarding implementation specifics.

\subsection*{Neural networks}
Neural networks (NN) are mathematical representation functions that came about via combination of multiple perceptrons \cite{bishop2006pattern}. A neural network-based classification system learns the features from the raw data and classifies the input data based on these features. They have recently gained popularity in image classification for their high accuracy. Here we review some of the main neural network techniques.

A popular neural network model is multilayer perceptron (MLP) \cite{gardner1998artificial}. A sequence of non-linear modules builds an MLP. Each module is comprised of a linear transformation unit followed by a nonlinear differentiable activation (generally ``ReLU'', ``sigmoid'', or ``tanh'') unit. Data samples, in their vectorized forms, enter the MLP from the input side and are successively transformed through all layers of MLP modules to generate outputs. Each layer of an MLP module transforms the input to a more abstract and higher-level representation. An interplay of these interconnected transformation layers enables the model to learn very complex functions. Due to considering input image data in vectorized forms, MLPs leave out the spatial coherence information of nearby pixels. Moreover, they are not always invariant to some transformations, e.g., translation, scaling, deformation, etc. that may be irrelevant in the context of some applications.

Convolutional neural network (CNN) is a class of neural networks \cite{simonyan2014very,szegedy2016rethinking, szegedy2015going} that is well suited for image data. Spatial coherence and transform invariance are built into the structure of CNNs. They are sensitive to relevant minute variations and invariant to irrelevant large variations, e.g., translation, scaling, etc. by considering local information in images. CNN layers have two different kinds of structures. The first few layers are comprised of a combination of convolutional, non-linear activation, and pooling layers. The remaining layers are denoted as fully connected layers that operate on data in their vectorized forms. A convolutional layer is subdivided into a number of small patches. Each patch extracts features from small sub-regions in images, filters those features, and passes the filtered features to the next layer. Each small patch in a particular layer shares the same sets of filter weights so that the same local pattern is detected at different locations of the image. Multiple features are detected with multiple convolutional layers. The pooling layers merge similar features together and reduce dimensionality to some extent. The lower-order information gathered in convolutional and pooling layers are combined with the higher-order features of fully connected layers later to compile the information about the image as a whole. The final decision layer is generally followed by a ``softmax'' function \cite{bishop2006pattern} that normalizes the outputs into numbers in the interval $(0,1)$.

A general architecture of a CNN is presented in  Fig.~\ref{figmethod}(b). An input image sample is processed by sub-units in convolutional layers with convolution and nonlinear mapping. The outputs of the convolutional layers are then passed to the pooling layers for semantic merging. The final layers of CNNs consist of fully connected layers. Multiple instances of convolutional, pooling, and fully connected layers are generally present in a standard CNN. A neural network-based classification system outputs a predicted class (e.g., benign, malignant, etc.) for an input image. Appendix A contains a more detailed mathematical description of neural networks.

Neural network-based cell image classification has been used to classify different staining patterns of human epithelial type-2 cells for detecting auto-antibodies of different autoimmune diseases \cite{gao2017hep,phan2016transfer,gao2014hep}, identify different subcellular localization patterns of proteins \cite{boland2001neural}, and diagnose cervical cancer \cite{yusoff2010performance}. Neural networks have outperformed many sophisticated artificial intelligence algorithms in the task of image classification but may require a large number of image samples to perform reliably \cite{gao2014hep,qi2016exploring}.

From the neural networks (NN) category, we compared the performances of a multilayer perceptron (MLP) and few convolutional neural networks (CNN): a shallow CNN implementation with one convolutional layer \cite{park2018multiplexing}, VGG16 (a deep neural network with 16 convolutional layers) \cite{simonyan2014very}, and Inception-V3 (also a deep neural network with 41 convolutional layers) \cite{szegedy2016rethinking, szegedy2015going}. VGG16 and Inception-V3 were implemented both without (VGG16, INCv3) and with (VGG16-T, INCv3-T) transfer learning using ``imagenet'' weights \cite{deng2009imagenet}. More information regarding implementation specifics can be found in Appendix B.

\subsection*{Transport-based morphometry}
Transport-based morphometry (TBM) methods, guided by the mathematics of optimal mass transport, decode differences among images by quantifying the least effort required to morph the images into a reference image \cite{kolouri2017,basu2014detecting,kolouri2016radon}. TMB methods transform raw image data into a representation that facilitates both image classification and visualization of biologically interpretable differences between classes. Embodiments of this nonlinear image representation method include Radon cumulative distribution transform \cite{kolouri2016radon}, continuous linear optimal transport \cite{kolouri2016continuous}, and discrete linear optimal transport \cite{basu2014detecting}, among others. 

The Wasserstein distance between two images can quantify the optimal transport of mass (image intensities, in the case of an image) required to morph one image into the other \cite{wang2013linear, basu2014detecting}. The weighted Euclidean distance between images in transport space is closely related to the Wasserstein distance between them in image space (refer to Basu~\emph{et al.}\cite{basu2014detecting}, Kolouri~\emph{et al.}\cite{kolouri2017optimal} for more details). TBM methods measure linear embeddings for all input images by computing their Wasserstein distances to a pre-computed reference image. This linear embedding produces the transport space representations of all the image samples. After image transformation, TBM employs a statistical regression-based classifier in transport space to separate data classes. One exclusive property of TBM methods is that it is possible to visualize and interpret any regression in transport space and thereby understand biologically interpretable differences across classes.

To understand the TBM mechanism, please refer to Fig.~\ref{figmethod}(c). Each point in the scatter plot in Fig.~\ref{figmethod}(c) embodies the transport space representation of an image sample in the subspace of the two most discriminant directions computed by penalized linear discriminant analysis (PLDA). A classifier is trained in the transport space to separate transport space representations of input images into different classes. Moreover, the differences among the classes are visualized by the inverse transform property of TBM. The representative images in the panels along the x and y-axes in Fig.~\ref{figmethod}(c) show the visualizations of class differences along the first and second most discriminant PLDA directions, respectively. Appendix A contains a more extensive description of TBM methods.

Transport-based morphometry has been used in a number of cell image classification problems: classification of liver hepatoblastoma by analyzing nuclear chromatin distribution \cite{basu2014detecting, wang2013linear}, classification of subcellular protein localization patterns in HeLa cells \cite{wang2013linear}, detection of follicular adenoma and carcinoma using histological images of thyroid nuclei \cite{wang2010detection, basu2014detecting}, and automated screening for cell phenotype changes as a result of drug treatment \cite{basu2014detecting}, to name a few.

From the transport-based morphometry (TBM) category, we implemented the Radon cumulative distribution transform (R-CDT) \cite{kolouri2016radon} coupled with the same statistical regression-based classifiers that were used with the numerical feature engineering category. Appendix B has more information regarding implementation specifics. For both the transport-based morphometry and the numerical feature engineering catagories, we did not implement SVM-k for the human epithelial cell dataset (Hep2) due to the computational complexity arising from a large dataset (63,445 images).

\section{Experimental Setup}
\label{secExp}
To test and compare the performances of three broad categories of cell image classification algorithms, we selected methods from a few exemplary papers: Wnd-chrm \cite{orlov2008wnd} from numerical feature engineering; multilayer perceptron \cite{gardner1998artificial} and three convolutional neural network architectures -- an existing shallow CNN \cite{park2018multiplexing}, VGG16 \cite{simonyan2014very}, and Inception-V3 \cite{szegedy2016rethinking, szegedy2015going} -- from neural networks; and the Radon cumulative distribution transform \cite{kolouri2016radon} from transport-based morphometry. Details of all these methods with mathematical descriptions are presented in appendix A.

\begin{figure*}
    \centering
    \includegraphics[width=17.5cm]{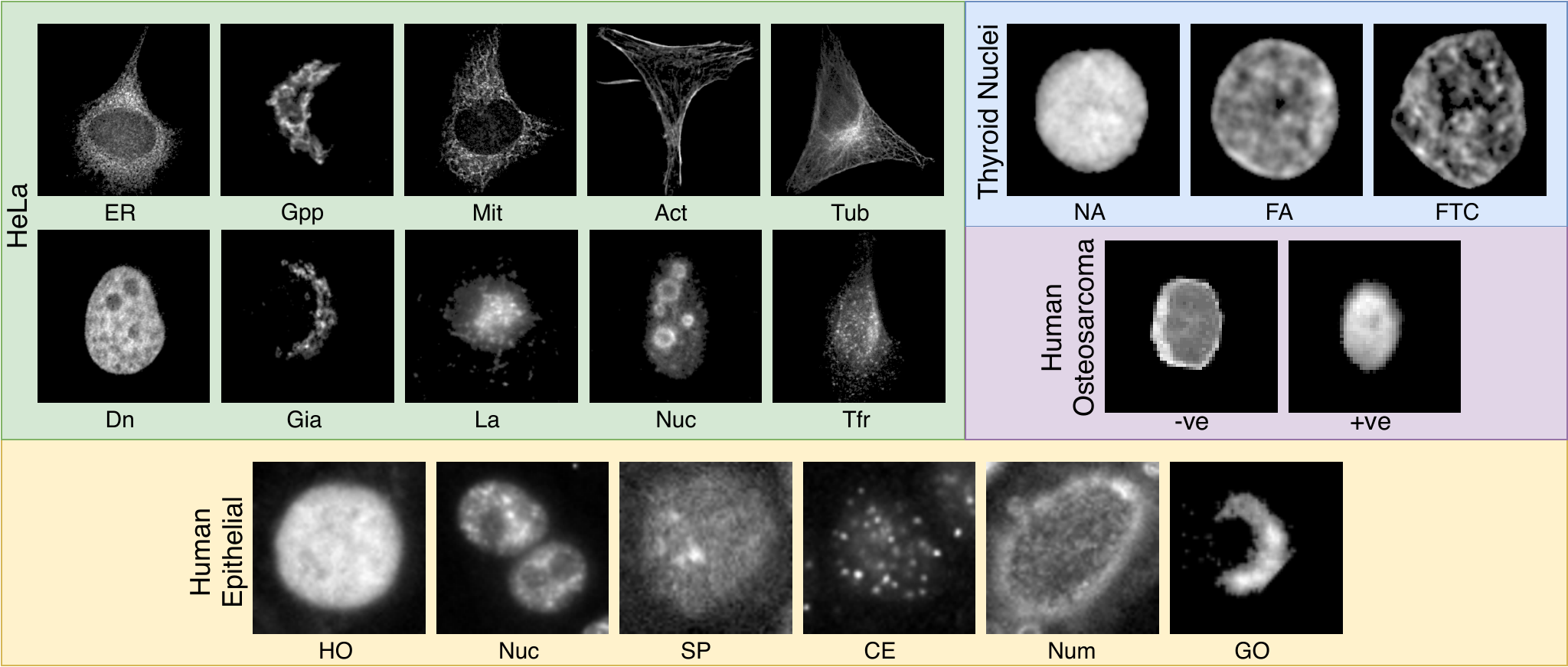}
    \caption{Datasets used for the cell image classification: subcellular protein localization patterns in HeLa cells; normal, benign, and malignant thyroid nuclei; human osteosarcoma cells before and after treatment with the drug, Wortmannin; and different staining patterns in human epithelial cells.}
    \label{figCells}
\end{figure*}
To evaluate the performance of each method, we trained each classification model using a portion (called training set) of a given dataset. The classification performance of the model was computed using the remaining portion (called the validation or test set) of the dataset. To reduce over-fitting or selection bias, the model performance was evaluated $k$ times using $k$ different partitions of training and testing sets via $k$-fold ($k=10$) cross-validation. All datasets were preprocessed as follows: all the images were centered such that the center of mass of each image occurs at the center of view of each image, oriented such that their major axes are aligned, and flipped such that they have similar intensity weight distribution (see Appendix A for details). Features were scaled to standardize the range of features using the standard scaling function in Python's scikit-learn package \cite{scikit-learn}. Principal component analysis was used to reduce data dimensionality.

\subsection*{Datasets}
Four different datasets with different classification problems were used in this overview: HeLa cell \cite{murphy2000towards,boland2001neural}, human osteosarcoma cell \cite{basu2014detecting,logan2010screening,carpenter2006cellprofiler}, thyroid nuclei \cite{basu2014detecting, wang2010detection}, and human epithelial cell \cite{qi2016exploring} image dataset. We used datasets where the cell images were already segmented. The classification performances of different methods were evaluated on each of these cell image datasets.
\subsubsection*{HeLa dataset}
The HeLa cell image dataset concerns a protein characterization problem in the field of functional genomics or proteomics: detecting subcellular protein localization patterns. The associated classification task is to separate major subcellular protein localization patterns to identify the distribution, and function of expressed proteins. Segmented fluorescence microscopy images of HeLa cells have been collected from Murphy~\emph{et al.}\cite{murphy2000towards} and Boland~\emph{et al.}\cite{boland2001neural} The HeLa cell image dataset contains 10 subcellular localization patterns of the major organelles: endoplasmic reticulum protein (ER), Golgi protein GPP130 (Gpp), mitochondrial protein (Mit), filamentous actin labeled with rhodamine-phalloidin (Act), cytoskeletal protein tubulin (Tub), DNA labeled with DAPI (Dn), Golgi protein giantin (Gia), lysosomal protein LAMP2 (La), nucleolar protein nucleolin (Nuc), and transferrin receptor (in endosomes) (Tfr) (see Fig. \ref{figCells}).
\subsubsection*{Human osteosarcoma cell dataset}
The Human osteosarcoma cell dataset \cite{basu2014detecting,logan2010screening,carpenter2006cellprofiler} is used for the task of examining the underlying trend of the cytoplasm-to-nucleus translocation of the forkhead fusion protein (FKHR-EGFP). The aim is to quantify the translocation of FKHR-EGFP with the infusion of Wortmannin dosage in stably transfected human osteosarcoma (U2OS) cells. Localized in the cytoplasm, FKHR usually moves towards the nucleus and then is transported out by export proteins. If this export is inhibited by some drug, e.g., Wortmannin, FKHR starts to accumulate in the nucleus. This accumulation may cause a cell phenotype change due to the export inhibition by Wortmannin. Images of human osteosarcoma cells have been collected from Carpenter~\emph{et al.}\cite{carpenter2006cellprofiler} and segmented using Basu~\emph{et al.}\cite{basu2014detecting}. In this overview, we have differentiated cells with no drugs (negative control) from cells with 150 nM Wortmannin added (positive control). Images from these two classes are illustrated in Fig. \ref{figCells}.

\subsubsection*{Thyroid nuclei dataset}
The thyroid nuclei dataset is used to distinguish normal, benign, and malignant cell types from nuclear structures of thyroid cells \cite{basu2014detecting, wang2010detection}. The task here is to distinguish among follicular adenoma (FA), follicular carcinoma (FTC), and normal (NA) thyroid cells (see Fig. \ref{figCells}) using only the information embedded in chromatin arrangements of the nuclei. FA and FTC are both neoplastic but only FTC is capable of metastases. Due to similarities in the nuclear features, differentiate between FA and FTC is difficult. Physicians have historically depended on histopathology to distinguish thyroid nuclei. After surgical removal, the lesion is examined for the capsular or vascular invasion, the characteristic feature of FTC. Changes in nuclear structures of thyroid cells is visible in microscopic images. Image analysis-based classification techniques can detect these changes. Segmented images of three classes of thyroid nuclei have been collected from Basu~\emph{et al.}\cite{basu2014detecting}.

\subsubsection*{Human epithelial cell dataset}
The dataset of human epithelial type-2 cells (Hep2) has been obtained from Qi~\emph{et al.}\cite{qi2016exploring}. The dataset contains a large number (63,445) of segmented cell images from six different classes. This dataset is created by segmenting out single cell images from 948 specimen images of the ICPR 2014 HEp-2 cell classification contest dataset. The classification task in this paper is differentiating various staining patterns in indirect immunofluorescence Hep2 images to indicate different antinuclear auto-antibodies related to autoimmune diseases. The six image classes in this dataset are centromere (CE), Golgi (GO), homogeneous (HO), nucleolar (NUC), nuclear membrane (NUM), and speckled (SP). Representative cell images from each image class are displayed in Fig. \ref{figCells}.

\subsection*{Evaluation Measures}
\subsubsection*{Percentage accuracy}
Each classification model was trained with a training set and evaluated with a testing set. We calculated the percentage accuracy of correctly predicting the class labels for test data in a given fold (or partition) of cross-validation. Let, $y_{i}^{(te)}$ be the percentage accuracy for a model for a testing set of the $i$-th fold. The mean, $\mu_{acc}$, and standard deviation, $\sigma_{acc}$, of the classification accuracy were calculated over $k$--folds of cross-validation as
\begin{eqnarray}
&\mu_{acc} = \frac{1}{k} \sum_{i=1}^{k}y_i^{(te)}\times 100\% \nonumber\\
&\sigma_{acc} = \sqrt{\frac{1}{k-1} \sum_{i=1}^{k} \left(y_i^{(te)}-\mu_{acc}\right)^2}\times 100\%\nonumber
\end{eqnarray}
\subsubsection*{Kappa statistic}
Cohen's kappa statistic, $\kappa$, measures the agreement among different realizations (partitions of cross-validation) of the algorithm by taking into account the agreement occurring by chance. $\kappa$ can be computed as
\begin{eqnarray}
\kappa = 1-\frac{1-p_0}{1-p_e} \nonumber
\end{eqnarray}
where $p_0$ is the relative observed accuracy and $p_e$ is the hypothetical probability of accuracy by chance (refer to Viera~\emph{et al.}\cite{viera2005understanding} and Landis~\emph{et al.}\cite{landis1977measurement} for more details).

\section{Results}
\label{secRes}
\begin{figure*}
    \centering
    \includegraphics[width=17.5cm]{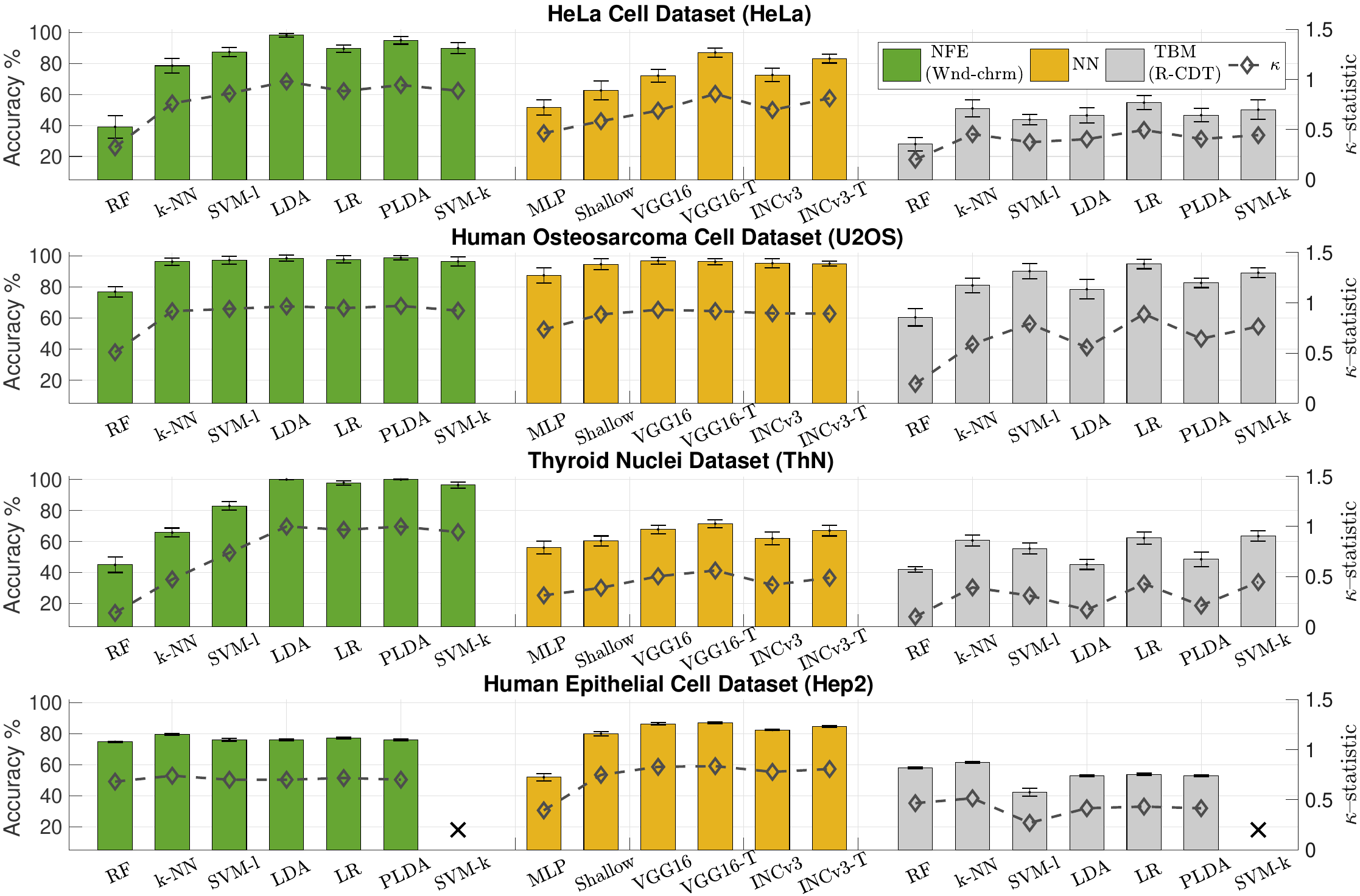}
    \caption{Percentage accuracy and $\kappa$-statistic values of the classification methods within the NFE, NN, and TBM-based categories in HeLa cell, human osteosarcoma cell, thyroid nuclei, and human epithelial cell image datasets.}
    \label{resfig01}
\end{figure*}

In this section, we present the comparative performances of different cell image classification techniques in terms of percentage accuracy and $\kappa$-statistic values. The classification techniques were divided into three broad categories: numerical feature engineering methods (NFE), neural network-based methods (NN), and transport-based morphometry methods (TBM). The methods were evaluated on four datasets: HeLa cell (HeLa), human osteosarcoma cell (U2OS), thyroid nuclei (ThN), and human epithelial cell (Hep2). A summary of all percentage accuracy and $\kappa$-statistic results computed in this paper is presented via bar graphs in Fig.~\ref{resfig01}. 

In the HeLa, U2OS, and ThN datasets, the classification accuracies within the NFE category (Wnd-chrm) are, in a majority of cases, higher than or equivalent to classification accuracies within the other two broad categories (Fig.~\ref{resfig01}). The performance superiority, in terms of classification accuracies, of the NFE-based category of methods for these three datasets is consistent across different classifier implementations, with a few exceptions, e.g., k-NN in HeLa and ThN, RF in HeLa, U2OS, and ThN, etc.

In contrast, for the Hep2 dataset, the classification accuracies of the majority of methods within the NN category are comparatively higher than the accuracies within the NFE and TBM-based categories of methods (Fig.~\ref{resfig01}). All of the CNN architectures implemented (Shallow, VGG16, VGG16-T, INCv3, INCv3-T) outperform, in terms of classification accuracies, all classifiers in each of the two other broad categories of methods.

The classification accuracies within the TBM category of methods (R-CDT) used in this paper vary across datasets and classifiers. Some classifiers implemented with the R-CDT produce results equivalent to a feature extraction method or a convolutional neural network. For example, in the U2OS dataset, R-CDT with LR (accuracy $=95\pm3.1$\%, $\kappa=0.89$) performs similarly to SVM-k with Wnd-chrm features (accuracy $=96\pm2.9$\%, $\kappa=0.92$), and in the ThN dataset, R-CDT with LR (accuracy $= 62\pm3.9$\%, $\kappa=0.43$) performs similarly to INCv3 CNN (accuracy $=62\pm4.2$\%, $\kappa=0.42$) (Fig.~\ref{resfig01}). In most of the cases, however, the TBM-based category of methods tested in this paper do not outperform NN and NFE-based methods.

\begin{figure}
    \centering
    \includegraphics[width=9cm]{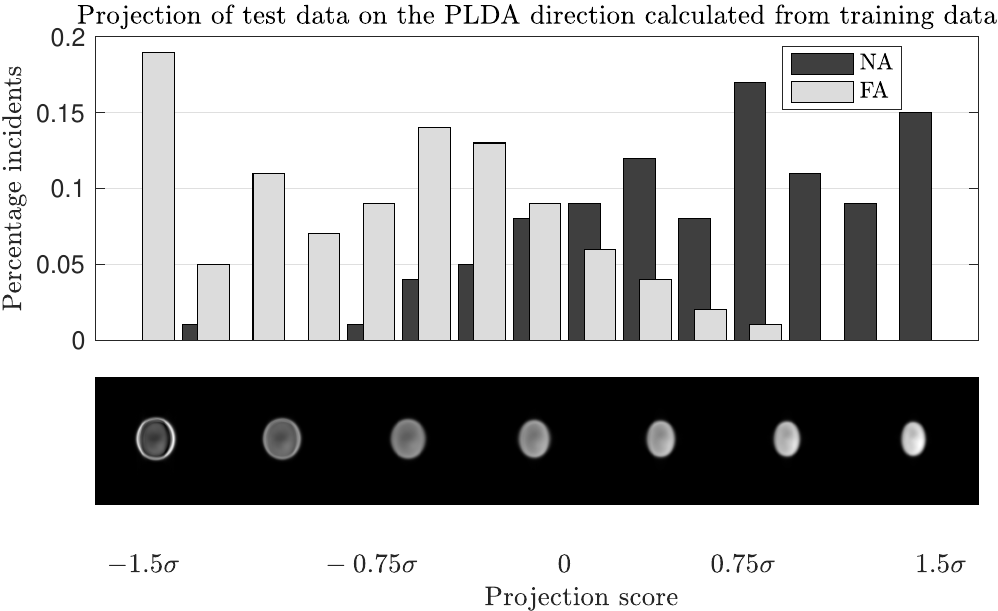}
    \caption{The histograms of the projections of the images of the normal (NA) and follicular adenoma (FA) subpopulations in the ThN dataset on the most discriminant PLDA direction. $\sigma$ denotes the standard deviation of the projection.}
    \label{resfig02}
\end{figure}

The ability of the TBM-based category of methods to visualize differences among the classes of a given dataset is presented in Fig.~\ref{resfig02}. We have chosen NA and FA classes of ThN dataset for demonstration. The histograms of the projections of R-CDT transformed test images along the most significant direction (computed using PLDA on training images) of differences between NA and FA cells are presented in the top panel, and the modes of variations of the images along the same discriminant direction is illustrated in the bottom panel in Fig.~\ref{resfig02}. The representative images at a particular coordinate along the discriminant direction correspond to the histograms at the same coordinate in the top panel. Important differences between the normal (NA) and follicular adenoma (FA) of the thyroid are evident in Fig.~\ref{resfig02}. It can be seen that peripheral chromatin concentration increases as the cell progresses from normal (NA) to follicular adenoma (FA) of the thyroid. The presence of peripheral rings and the decline of homogeneity of chromatin concentration are revealed by the TBM mechanism to be the possible underlying differences between follicular adenoma (FA) and 
normal (NA) thyroid tissue.

\begin{figure}
    \centering
    \includegraphics[width=9cm]{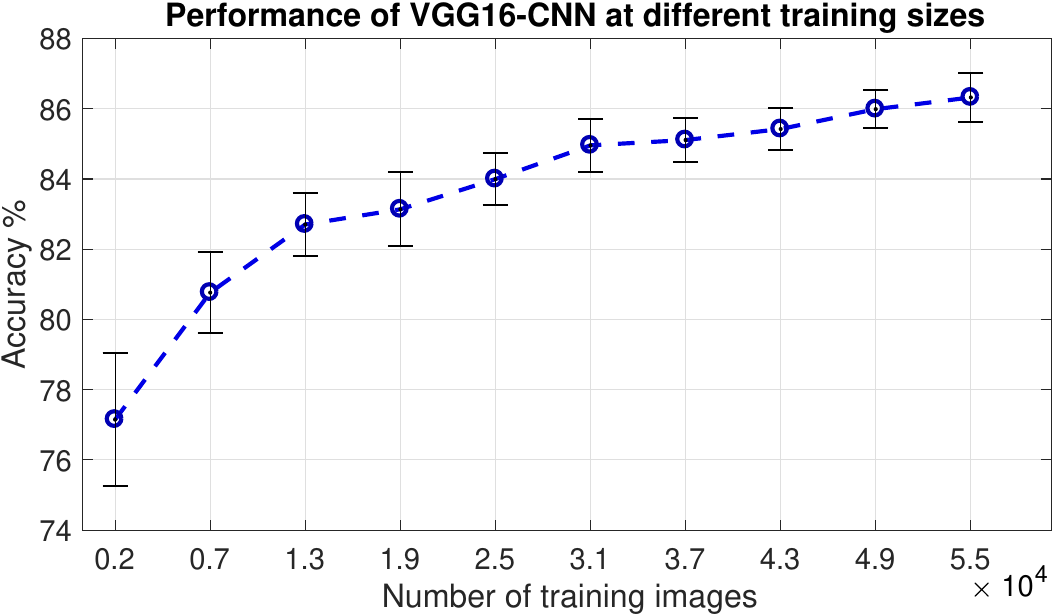}
    \caption{Percentage accuracy of VGG16-CNN in Hep2 dataset for different sizes of the training data.}
    \label{resfig03}
\end{figure}

To understand the impact of the dataset size on the classification accuracy of neural networks, we measured the percentage accuracy of VGG16-CNN for different training dataset sizes. To that end, we selected the dataset with the largest number of data samples (Hep2 -- 63,445 images). We sub-sampled Hep2 dataset into sub-datasets with training dataset sizes ranging from $2000$ to $55,000$ and evaluated the performance of VGG16-CNN on each subset with $10$-fold cross-validation. In this experiment, the percentage accuracy of VGG16-CNN increased as the number of training images increased (Fig.~\ref{resfig03}).

The values of the standard deviation of the accuracy for all the methods are in an acceptable range in all four datasets. This attests to the reliability of the estimates. Also, the Cohen's $\kappa$ statistic, that measures of agreements among different realizations, is $>0.6$ in most of the cases (indicating substantial or almost perfect agreement according to Landis~\emph{et al.}\cite{landis1977measurement}), in the range $(0.2,~0.4)$ in many cases (indicating fair or moderate agreement according to Landis~\emph{et al.}\cite{landis1977measurement}), and $<0.2$ in few cases (indicating slight agreement according to Landis~\emph{et al.}\cite{landis1977measurement}). The substantial or fair agreements as quantified by the Cohen's $\kappa$ statistic in most of the cases indicate equivalent partitions among data classes and pronounces robustness of the evaluation measures.

\section{Discussions and Conclusion}
\label{secDiscnConcl}
This paper has presented a comparative overview of different cell image classification techniques. Currently available methods for cell image classification were divided into three basic categories: numerical feature engineering (Wnd-chrm), neural networks (MLP, CNN), and transport-based morphometry (R-CDT). The methods were tested and compared in a standardized manner on four different cell image datasets with different classification problems: subcellular protein localization from microscopy images \cite{murphy2000towards,boland2001neural}, the quantification of the FKHR-EGFP translocation variation due to the addition of a drug \cite{basu2014detecting,logan2010screening,carpenter2006cellprofiler}, differentiation between benign and malignant thyroid nuclei \cite{basu2014detecting, wang2010detection}, and classification of human epithelial datasets based on cell staining patterns \cite{qi2016exploring}. 

\subsection*{Which method does have the best accuracy?}
The percentage accuracy of the numerical feature-based method, Wnd-chrm is the best or near the best in all four datasets. Wnd-chrm constructs a comprehensive feature vector with a large set of empirical representations (1025 features) of each image \cite{orlov2008wnd}. Though the extracted features, as well as their linear or nonlinear combinations used in classifiers, lack intuitive explanation regarding the underlying biological mechanisms, for the problems we tested, they exhibit high effectiveness in the context of cell image classification. Among the statistical regression-based classifiers that have been used with the Wnd-chrm features, the percentage accuracies of LR, LDA, SVM-k, and PLDA are consistently higher across all datasets.

As far as the neural networks category of methods, we compared two kinds: multi-layer perceptron (MLP), and convolutional neural networks (CNN). The MLP method we tried consistently underperformed other methods. The CNN methods we tried produced state of the art results only in classifying different staining patterns in Hep2 cell images but not in other classification problems. Unlike other datasets, the Hep2 cell image dataset contains a large number of image samples -- 63,445 images (approximately 10,570 images per class). Other datasets do not have such a big number of image samples: HeLa dataset contains 862 images (approximately 86 images per class), U2OS dataset contains 492 images (approximately 246 images per class), and ThN dataset contains 2053 images (approximately 684 images per class). Thus, we hypothesize that a large number of training samples may be necessary for neural networks to perform better.

The transport-based morphometry method, using the R-CDT, underperformed other methods in terms of percentage accuracy. However, R-CDT can potentially help to explain biological processes by providing visualizable models for mass distribution in cell images (see Fig.~\ref{resfig02}) and thus elucidate underlying cell mechanisms and provide with the opportunity for hypothesis generation.

\subsection*{How much data is enough for CNNs?}
CNN-based methods have been used in many cell biology and digital pathology applications \cite{qi2016exploring,gao2017hep,phan2016transfer,gao2014hep,boland2001neural} where dataset sizes range from $\sim800$ to $\sim6\times10^4$. To understand the impact of dataset sizes on the performance of neural networks, we sub-sampled the largest dataset, Hep2 (63,445 image samples) into few smaller datasets (with smaller training and testing datasets) and evaluated the performance of VGG16-CNN on each of them with a $10$-fold cross-validation. We did not conduct this experiment on other datasets because they do not contain such a large number of image samples as the Hep2 dataset. The plot of the percentage accuracy of VGG16-CNN in Hep2 dataset for different training dataset sizes is illustrated in Fig.~\ref{resfig03}. It can be seen that the VGG16-CNN outperforms Wnd-chrm with k-NN (accuracy = $79.39\pm0.60\%$) when the training dataset size $\geq\sim7,000$. However, this lower bound cannot be used as a guideline for training dataset sizes for CNNs as their performances depend on both the number of training samples and the complexity of the problem.

Different applications in cell biology and medicine employ different numbers of cell images. The size of the datasets in some drug discovery applications \cite{logan2010screening,carpenter2006cellprofiler,perlman2004multidimensional} ranges from $\sim500$ to $\sim7\times10^7$. The number of data samples varies from $\sim250$ to $\sim6\times10^4$ in many digital pathology applications \cite{ponomarev2014ana, yang2018, qi2016exploring, theera2007morphological,ko2011cell, theera2005white}. Some cell biology applications \cite{murphy2000towards, boland2001neural} have dataset sizes of the order $\sim800$. Given the perceived lack of theory about CNNs, we cannot recommend a specific number of training images for CNNs for a specific application. But as is apparent from the experiment in Fig.~\ref{resfig03}, greater numbers of training images lead to better CNN performance, as measured by classification accuracy.

\subsection*{Does CNN performance depend on architectures?}

In order to assess the performance of CNN applications to cellular classification tasks, we utilized two different frameworks, VGG-16 \cite{simonyan2014very}, developed in 2014, and Inception-V3 \cite{szegedy2016rethinking, szegedy2015going}, developed in 2016. Based on some experiments \cite{szegedy2016rethinking}, it may be reasonable to expect that Inception-V3 would outperform VGG-16 for the same task. Instead, what we observe is, in some cases, the classification accuracy of Inception-V3 is lower compared with that of VGG-16 (e.g., ThN, Hep2 datasets). In addition to evaluating two frameworks, we investigated whether transfer learning \cite{pan2010survey}, a technique widely adopted in some problems \cite{raina2006constructing, dai2007transferring, blitzer2006domain, wu2004improving, zheng2008transferring}, would improve classification accuracies as well. In our classification problems, transfer learning improved the performance of CNNs to some extent, especially for datasets with a relatively lower number of samples (e.g., HeLa, ThN), but still did not outperform numerical feature engineering methods. Overall, our experimental results suggest that the choice of architecture may not significantly alter classification results.

\subsection*{Which method is the easiest to use?}
A practical concern for researchers looking to perform cell image classification is which available method has the greatest ease of use. While considerations such as data size/type and theoretical reasoning are critical concerns, research is frequently constrained by realistic limitations such as time, computing power, and user expertise.

One important aspect to consider when choosing among various cell classification methods might be the time available for computation. Time concerns may depend on the type of question being asked and the associated urgency with which an answer is required. For example, in clinical use, determining whether cells from a tissue sample are benign or malignant is time-sensitive due to the role that the outcome plays in a patient's course of treatment. In this case, a method with a computation time of multiple days may not be suitable. For other questions, however, such as general scientific inquiries, methods which require more time may be acceptable. In a similar vein, computation time is often associated with the amount of training data available. As noted previously, CNNs perform better as training data set size increases, but this increase in accuracy comes with an increase in computational time as well. The trade-off between incremental increases in accuracy and a rise in computation time should be assessed on a case-by-case basis.

In addition to time constraints, computational resources affect which classification methods may be used. Wnd-chrm and R-CDT calculations can be performed on low-level machines whereas CNN calculations require much higher computing power. This additional requirement may limit when CNNs can be executed. 

An additional point of consideration when determining which classification method is most suitable for an application is the domain-specific expertise required to execute each method. Executing both feature extraction via Wnd-chrm and the R-CDT computation are a simple matter of inputting the correct dataset path and running the code, with no intermediate steps. The straightforward nature of these methods makes them suitable for settings in which individuals have little to no prior experience with classification techniques. CNNs, on the other hand, require additional steps beyond identifying the dataset path. Manual tuning of the batch size, learning rate, and step-size is a necessary step to achieving optimal accuracy. Adjusting these parameters appropriately requires some knowledge of both general coding techniques and machine learning theory, making this method suitable for use only in environments in which an individual has prior knowledge and experience with CNNs.

\subsection*{Which method should I use for my problem?}
Beyond time, computing power, and expertise constraints, another important factor to consider in cell classification method selection is which method is best suited for the nature of the problem. If we are interested in understanding the underlying cell biology, the transport-based morphometry methods can provide an visual representation of any regression methods. If we are interested primarily in achieving high classification accuracy, either feature engineering or neural network-based approaches are the preferred methods. In terms of classification accuracy, the numerical feature engineering methods may be suitable choices for cell image classification problems because they have the best or near the best accuracies in the datasets we tested. Neural network-based methods may be feasible when the number of training samples is very high (approximately on the order of 7,000 images or more for Hep2 dataset). In many cell image classification applications, the number of images available may be low, and, therefore, neural network-based methods may not be optimal. Because selecting a method for cell image classification tasks is problem-specific, we have provided a Python code that implements each method.    




\section*{Acknowledgements}
This work was supported in part by National Institutes of Health awards GM130825 and GM090033.

\section*{Source code}
\textbf{Source code} is available at \url{https://github.com/rohdelab/cell-image-classification}.\\

\section*{Appendix A. Mathematical Details of Image Classification Methods}
\subsection*{Image Preprocessing}
All images underwent several preprocessing steps. We first converted all of the images to grayscale. Because we were interested in the class differences that were independent of rigid body transformations, e.g.,  rotations, translations, reflections, etc., we initialized the images after minimizing the following functional: 
\begin{align}
&\Psi\left(\mathbf{A}_1,...,\mathbf{A}_N,\mathbf{r}_1,...,\mathbf{r}_N\right)\nonumber\\&=\sum_{m=1}^{N-1}\sum_{n=m+1}^{N}\int_{\Omega}\left|I_m^o\left(\mathbf{A}_m\mathbf{x}+\mathbf{r}_m\right)-I_n^o\left(\mathbf{A}_n\mathbf{x}+\mathbf{r}_n\right)\right|^2 dx\nonumber
\end{align}
where, $I_m^o(\mathbf{x})$ is the vectorized form of the $m$-th sample of original raw images, $\mathbf{A}_m$ is a matrix parameterized by rotation and isotropic scaling, and $\mathbf{r}_m$ is the translation vector. The minimization of such functionals is computationally intensive. Therefore, to yield the preprocessed image, $I_m^p(\mathbf{x})$, the following approximation was applied to the original raw image, $I_m^o(\mathbf{x})$, as in Rohde~\emph{et al.}\cite{rohde2008deformation}: the center of mass of each image was translated to the center of view of each image, the principal axis of each image was aligned to a predetermined angle, and the images were flipped to have similar intensity weight distribution by switching co-ordinates until the functional was minimized.

\subsection*{Numerical feature engineering -- Wnd-chrm}
In numerical feature engineering, predetermined features are extracted from cell images, and those features are used to classify cells. For example, a vector of extracted features can be formulated for a particular image as
\begin{align}
    F\left(~\vcenter{\hbox{\includegraphics[width=0.75cm]{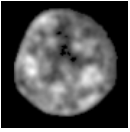}}}~\right)=
    \begin{bmatrix}
    F_1=\text{area}\\
    F_2=\text{perimeter}\\
    F_3=\text{Harlick texture}\\
    \vdots\\
    F_N=\text{Euler number}
    \end{bmatrix}
    \nonumber
\end{align}
The process of feature vector representation is repeated for all image samples. Once feature extraction is completed for all samples, a classifier is trained in this feature vector space to classify cell images. We selected Wnd-chrm \cite{orlov2008wnd} as the feature extraction program from the numerical feature engineering category for cell image classification. Although Wnd-chrm has built-in methods for classification, we used Wnd-chrm exclusively as a feature extraction method. Using Wnd-chrm features, the feature vector, $I_m^f(\mathbf{x})$, for the $m$-th image sample was constructed from the corresponding preprocessed image, $I^p_m(\mathbf{x})$. Wnd-chrm extracts a large set of 1025 image descriptors including high contrast features (e.g., Gabor textures, edge statistics, object statistics, etc), polynomial decompositions (Chebyshev statistics, Zernike polynomials, etc), and statistics and textural information (first four moments, multiscale histogram, Haralick textures, Tamura textures, Radon transform statistics, etc) \cite{orlov2008wnd}. These features are calculated on raw pixel images and transforms of the images (wavelets, Fourier, Chebyshev transforms), as well as the transforms of the image transforms. For more details of Wnd-chrm-based feature extraction, refer to Orlov~\emph{et al.}\cite{orlov2008wnd}.

\subsection*{Neural networks -- MLP, CNNs}
Multilayer perceptron (MLP), a class of feedforward neural network architecture, is given as a series of nonlinear modules containing a linear transformation unit followed by a non-linear activation unit. The MLP-based classification model for a $P$-class classification problem is
\begin{align}
    \sigma\left(\Theta_L\sigma\left(\cdots\sigma\left(\Theta_2\sigma\left(\Theta_1\mathbf{v}+\mathbf{b}_1\right)+\mathbf{b}_2\right)+\cdots\right)+\mathbf{b}_L\right)=
    \begin{bmatrix}
    \text{OTN-1}\\
    \text{OTN-2}\\
    \text{OTN-3}\\
    \vdots\\
    \text{OTN-P}
    \end{bmatrix}\nonumber
\end{align}
where $\mathbf{v}$ is the input image in the vectorized form, $\Theta$ is the weight matrix, and $\sigma:\mathbb{R}^n\to\mathbb{R}^n$ is an activation function. OTN-P denotes the P-th output node of MLP that indicates the predicted class for an input image. In our work, the network parameters were trained with all the preprocessed training images, i.e., $\mathbf{v}=$ the vectorized form of the $m$-th preprocessed training image,  $I_m^{p}(\mathbf{x}),~\forall m$. The gradient of the energy function was computed using the backpropagation algorithm \cite{goodfellow2016deep}.

Convolutional neural networks (CNN) have different structures in the initial modules (or layers). The first few layers of a CNN are comprised of convolutional, non-linear activation, and pooling layers. A convolutional layer extracts and filters features from small sub-regions in images and passes the filtered features to the next layer. The sub-regions in a particular layer share the same sets of filter weights. The pooling layers merge semantically similar features together. The rest of the layers are fully connected and operate on data in their vectorized forms.

\subsection*{Transport-based mophometry -- R-CDT}
From the transport-based morphometry category, a nonlinear and invertible image transform method known as the Radon cumulative distribution transform (R-CDT) \cite{kolouri2016radon} was selected. We selected R-CDT because of requiring less computation and producing closed form solutions without resorting to any kind of optimization algorithm. Before a classifier was run, each preprocessed image, $I^p_m(x,y)$, was converted to its corresponding R-CDT representation. The procedure for calculating the R-CDT of an image had several steps. A template image, $I^p_0(x,y)$, was computed as the mean of all of the $M$ preprocessed images, $I^p_m(x,y),~\forall m\in[M]$. The image samples and the template image were normalized such that
\begin{eqnarray}
\int_{\mathbb{R}^2}I^p_m(x,y)dxdy=\int_{\mathbb{R}^2}I^p_0(x,y)dxdy=1\nonumber
\end{eqnarray}
After normalization, $I_m^p$ and $I_0^p$ can be treated as probability density functions. Next, sinograms were obtained from the Radon transforms of the template and all the sample images as follows:
\begin{eqnarray}
\hat{I}_m^p=\mathcal{R}\left(I_m^p\right);\hat{I}^p_0=\mathcal{R}\left(I_0^p\right)\nonumber
\end{eqnarray}
where
\begin{align}
\hat{I}(t,\theta)&=\mathcal{R}\left(I(x,y)\right)\nonumber\\
&=\int_{-\infty}^{\infty}\int_{-\infty}^{\infty}I(x,y)\delta\left(t-x\mbox{cos}(\theta)-y\mbox{sin}(\theta)\right)dxdy\nonumber
\end{align}
Provided that $\mu_m$ and $\sigma$ are continuous probability measures on $\mathbb{R}^2$ with corresponding probability density functions $I_m^p$ and $I^p_0$, for a fixed angle $\theta$, there exists a unique one-dimensional measure preserving map, $f_m(.,\theta)$ that warps $\hat{I}^p_m(.,\theta)$ into $\hat{I}_0(.,\theta)$ satisfying the following:
\begin{align}
\int_{-\infty}^{f_m(t,\theta)}\hat{I}^p_m(\tau,\theta)d\tau=\int_{-\infty}^{t}\hat{I}^p_0(\tau,\theta)d\tau,\hspace{0.4cm}\forall\theta \in [0,\pi]\nonumber
\end{align}
The forward R-CDT for a sample image, $I_m(x,y)$, is then defined as
\begin{eqnarray}
I_m^t(.,\theta)=\left(f_m(.,\theta)-id\right)\sqrt{\hat{I}^p_0(.,\theta)}\nonumber
\end{eqnarray}
where, $id:\mathbb{R}\rightarrow\mathbb{R}$ is the identity function, $id(x)=x$. After R-CDT computations were complete for all images, a classifier was trained in transform space to separate the cell images.

\subsection*{Statistical regression-based classifiers}
Statistical regression-based classifiers estimate linear or non-linear decision boundaries in the feature or transport space to classify images. In this paper, a number of classifiers were tested: linear discriminant analysis (LDA), penalized linear discriminant analysis (PLDA), support vector machine (with both a linear and radial basis function kernel: SVM-l and SVM-k, respectively), logistic regression (LR), random forests (RF), and k-nearest neighbors (k-NN). Most of the classifiers were used with the default settings of the scikit-learn \cite{scikit-learn} package in Python.

The vectorized forms of the R-CDT transform domain images, $I_m^t$, and the Wnd-chrm feature images, $I_m^f$ (let us drop the indexing with `$\mathbf{x}$' or `$(x,y)$'), were analyzed by statistical regression-based classifiers. We denote both $I_m^t$ and $I_m^f$ as $I_m$ in this section. The dataset was split into training and testing sets, $I_m^{(tr)}$ and $I_m^{(te)}$, respectively, by stratified 10-fold cross-validation. The classifiers were trained on the training data and their performances were evaluated using the testing data. A brief description of the classifiers used are presented in the following:
\subsubsection*{Linear Discriminant Analysis (LDA)}
Linear discriminant analysis classifier differentiates data classes with a linear combination of features based on Fisher's linear discriminant \cite{fisher1936use, mclachlan2004discriminant}. LDA classifiers estimate a linear decision boundary by maximizing the following objective function with respect to the matrix of discriminant directions:
\begin{eqnarray}
J(\mathbf{W})=\frac{\mathbf{W}^T\mathbf{S}_b\mathbf{W}}{\mathbf{W}^T\mathbf{S}_w\mathbf{W}}\nonumber
\end{eqnarray}
where, $\mathbf{S}_b=\sum_{i=1}^cN_i^{(tr)}(\mu_i-\mu)(\mu_i-\mu)^T$ with $\mu=\frac{1}{N^{(tr)}}\sum_{\forall m\in[N^{(tr)}]}I_m^{(tr)}$,  $\mu_i=\frac{1}{N^{(tr)}_i}\sum_{I_m^{(tr)}\in \mathbf{w}_i}I_m^{(tr)}$, $N^{(tr)}$ = number of total training data samples, $N^{(tr)}_i$ = number of training data samples in the $i$-th class, $\mathbf{S}_w=\sum_{i=1}^c\sum_{I_m^{(tr)}\in \mathbf{w}_i}(I_m^{(tr)}-\mu_i)(I^{(tr)}_m-\mu_i)^T$, $c$ is the number of linear discriminant directions, and $\mathbf{W}$ is a matrix formed by concatenating the discriminant directions, $\mathbf{w}_i,~i\in[c]$, in its columns. In addition to Fisher's linear discriminant analysis (LDA) classifier, we also used a penalized version of linear discriminant analysis (PLDA) classifier \cite{hastie1995penalized}.
\subsubsection*{Support Vector Machine (SVM)}
Support vector machine or the maximum margin classifier \cite{suykens1999least} aims to separate
data classes with a hyperplane such that the separating margin between different classes is maximized. The hyperplane is estimated by minimizing the following
constrained objective function:
\begin{align}
    &\min_{\mathbf{w}\in\mathbb{R}^n,~\mathbf{s}\in\mathbb{R}^{N^{(tr)}}}\frac{1}{N^{(tr)}}\sum_{m=1}^{N^{(tr)}}s_m+\frac{\lambda}{2}||\mathbf{w}||^2_2 \nonumber\\
    &\mbox{subject to}~s_m\geq0,~y_{m}^{(tr)}\left(\mathbf{w}^TI_m^{(tr)}\right)\geq1-s_m~\forall m\in[N^{(tr)}]\nonumber
\end{align}
where $\mathbf{w}$ is the SVM hyperplane. A nonlinear version of SVM can also be constructed by incorporating a kernel function in the dual formulation of the SVM optimization problem \cite{amari1999improving}. We implemented both the linear (SVM-l) and the kernel (SVM-k) versions in this paper. The SVM classifiers were run with the default settings in the scikit-learn package \cite{scikit-learn}.
\subsubsection*{Logistic Regression (LR)}
The logistic regression classifier \cite{kleinbaum2002logistic} -- a form of binomial regression -- uses a logistic or ``sigmoid'' function to identify the differences in data patterns. LR classifiers estimate a linear decision boundary by minimizing the following negative log-likelihood function:
\begin{align}
	    \min_\mathbf{w}~-&\sum_{m=1}^{N^{(tr)}}(y_m^{(tr)}\log p(y=y_m^{(tr)}|I_m^{(tr)},\mathbf{w})\nonumber\\&+(1-y_m^{(tr)})\log(1- p(y=y_m^{(tr)}|I_m^{(tr)},\mathbf{w})))\nonumber
\end{align}
where
\begin{align}
	    &p(y=1|\mathbf{x},\mathbf{w})=g\left(\mathbf{w}^T\mathbf{x}\right)=\frac{1}{1+e^{-\mathbf{w}^T\mathbf{x}}},\nonumber\\~~~&p(y=0|\mathbf{x},\mathbf{w})=1-p(y=1|\mathbf{x},\mathbf{w}),\nonumber
\end{align}
and $\mathbf{w}$ is the linear decision boundary of the LR classifier. The default settings in the scikit-learn package \cite{scikit-learn} were used for the LR classifier.
\subsubsection*{Random Forests (RF)}
The random forests classifier \cite{ho1995random}, an ensemble learning algorithm, constructs multiple decision trees and classifies based on the mean or the mode of the prediction results of individual trees. A random forests classifier repeatedly selects random subsets of data with replacement from the training samples, $I_m^{(tr)}$, and trains a decision tree classifier on each of the random subsets. After training, the prediction for a testing data sample, $I_m^{(te)}$, can be made either by averaging or by taking the majority vote of the predictions from all the individual decision trees. We implemented RF classifiers with the default settings in the scikit-learn package \cite{scikit-learn}.
\subsubsection*{$k$--Nearest Neighbors (k-NN)}
The $k$--nearest neighbors classifier assigns classes to an unknown data point (cell images, in our case) by taking majority votes of the $k$ nearest training data points around it. For an unknown data point, $I_m^{(te)}\in \mathbb{R}^d$, the classifier forms a set $\mathcal{A}$ with the $k$ nearest (in terms of Euclidean distance) training points. Then the conditional probability of the data point $I_m^{(te)}$ to belong to the class $j \in \{0,1,\cdots,n\}$ is estimated as
\begin{eqnarray}
P(y=j|\mathbf{X}=I_m^{(te)})=\frac{1}{k}\sum_{i\in\mathcal{A}}J(y^{(tr)}_{i}=j)\nonumber
\end{eqnarray}
where $J(x)$ is the indicator function which is $1$ if $x$ is $true$, and $0$ otherwise. After the conditional probability has been estimated, $I_m^{(te)}$ is assigned to the class with the largest probability.In our paper, the k-NN classifier was used with the default settings in the scikit-learn package \cite{scikit-learn}.

\section*{Appendix B. Cell Image Classification Software}
We have provided a python code for implementing all the cell image classification methods reviewed in this paper. The source code for our experiments is available at \url{https://github.com/rohdelab/cell-image-classification}. The usage instructions for the software are as follows:

\subsection*{What packages have to be installled}
To use the software, first install the following dependences: Python 3.6, Tensorflow 1.13.1, scikit-learn 0.18.1, wnd-charm and its Python API (\url{https://github.com/wnd-charm/wnd-charm}), and the Python optimal transport library (\url{https://github.com/LiamCattell/optimaltransport}).

\subsection*{How data should be organized}
To run the experiments on a cell image dataset, first create a directory under the `data' directory of the downloaded folder. Then place the segmented and preprocessed images of different classes in different sub-directories of the newly created directory. As an example, the HeLa cell image dataset is provided with the Python software following the required organization.

\subsection*{What commands have to be run}
After dependencies are installed and data are placed in correct directories, we can run various classification methods in command prompt. For example, to run logistic regression using Wnd-chrm  features run the following command in the top directory containing the downloaded code: ``python main.py - - dataset example\_directory - - space wndchrm - - model LR''. The options for `- - space' include `image', `wndchrm', and `RCDT', while for `- - model' include `RF', `KNN', `SVM', `LR', `LDA', `PLDA', `MLP', `ShallowCNN', `VGG16', and `InceptionV3'. For detailed instructions please refer to the README file included in the github repository.

\bibliography{refs}

\end{document}